\documentstyle[preprint,aps,12pt]{revtex}
\tightenlines
\begin{document}
\draft
\hfill\vbox{\baselineskip14pt
             \hbox{\bf ETL-00-100}
            \hbox{March 2000}}
\baselineskip20pt
\vskip 0.2cm 
\begin{center}
{\Large\bf Parameters for Systems Exhibiting Local Lattice Distortions,
Charge and Spin Ordering}
\end{center} 
\vskip 0.2cm 
\begin{center}
\large Sher~Alam,~T.~ Yanagisawa, and H. Oyanagi
\end{center}
\begin{center}
\end{center}
\begin{center}
{\it Physical Science Division, ETL, Tsukuba, Ibaraki 305, Japan}
\end{center}
\vskip 0.2cm 
\begin{center} 
\large Abstract
\end{center}
\begin{center}
\begin{minipage}{14cm}
\baselineskip=18pt
\noindent

	Keeping in mind the experimental results that indicate 
local lattice distortions, charge and spin orderings, we have developed a 
phenomenological approach which allows us to describe the electronic
phase diagram of cuprates and related systems in terms of few
parameters.

In the present work we consider a third-order parameter theory which
characterize charge, spin and superconductivity orderings. We are
thus led to a theory of three scalar fields. By coupling these
scalars to gauge fields we are naturally led to string-like
solutions, which we interpret as stripes. This ties nicely
with our quantum group conjecture that 1d systems play
an important role in the physics of cuprates and related
materials.
 
We show that this simple approach can give rough values
for two-order parameters which can be naively be interpreted
as charge and spin orderings. We also report our attempt to 
understand how local lattice distortions are involved and what 
role they play in terms of these two order parameters.

\end{minipage}
\end{center}
\vfill
\baselineskip=20pt
\normalsize
\newpage
\setcounter{page}{2}
\section{Introduction}
	In a previous work one of us \cite{alam98} has advanced 
the conjecture that one should attempt to model the phenomena of
antiferromagnetism and superconductivity by using quantum
symmetry group. Following this conjecture to model the phenomenona of
antiferromagnetism and superconductivity by quantum symmetry
groups, three toy models were proposed \cite{alam99-1}, namely,
one based on ${\rm SO_{q}(3)}$ the other two constructed with
the ${\rm SO_{q}(4)}$ and ${\rm SO_{q}(5)}$ quantum groups. 
Possible motivations and rationale for these choices  
were outlined \cite{alam99-1}. In \cite{alam99-2} a model to 
describe quantum liquids in transition from 1d to 2d dimensional 
crossover using quantum groups was outlined. In \cite{alam00-1} 
the classical group ${\rm SO(7)}$ was proposed as a toy model to 
understand the connections between the competing phases and the 
phenomenon of psuedo-gap in High Temperature Superconducting Materials
[HTSC]. Then we proposed in \cite{alam00-2} an idea to
construct a theory based on patching critical points so as to
simulate the behavior of systems such as cuprates.
To illustrate our idea we considered an example 
discussed by Frahm et al., \cite{fra98}. The model
deals with antiferromagnetic spin-1 chain doped with
spin-1/2 carriers. In \cite{alam00-3} the connection between
Quantum Groups and 1-dimensional [1-d] structures such as
stripes was outlined. The main point of \cite{alam00-3}
is to emphasize that {\em 1-d structures play an important
role in determining the physical behaviour [such as the 
phases and types of phases these materials are capable of
exhibiting] of cuprates} and related materials.

	In this note we examine a phenomenological and
classical model to understand the phenomenon of ``ordering''
in HTSC and related materials.

\section{Symmetry Breaking}
	It is well-known that the Higgs-model in particle
physics has its origins in condensed matter physics [CMP] and
that Higgs-model is a model of spontaneous symmetry
breaking [SSB]. A well-known example of SSB in CMP
is ferromagnetism near the Curie temperature. When
the temperature is greater than Curie temperature
all dipoles are randomly oriented and the ground-state
is {\em rotationally} invariant, whereas when the
temperature is below the Curie temperature all the
dipoles are aligned along an arbitrary direction
giving rise to spontaneous magnetization and
rotational symmetry is {\em hidden}. In other words
we can say that we have SSB when the symmetry of
Hamiltonian or Lagrangian is not explicitly
shared by the ground-state or the vacuum. 
Thus the symmetry breaking condition is the
non-invariance of the vacuum (ground-state)
\begin{eqnarray}
U |0> \neq |0> ,
\label{s0}
\end{eqnarray}
where $U$ is an element of the symmetry group.

Ginzburg-Landau theory [GL] is a phenomenological
model for mainly second-order phase transitions.
The main-point of GL can be easily understood by
resorting to ferromagnetism as an example.
The essential point is that for temperatures
near the Curie temperature, the magnetization 
{\bf M} is assumed to be small and one can
Taylor expand the free energy density as 
\begin{eqnarray}
f({\bf M}) &=& (\partial_{i} {\bf M})^{2}
+V({\bf M}),\nonumber\\
 V({\bf M}) &=& \alpha_{1}(T)({\bf M}\cdot {\bf M})
+\alpha_{2}(T)({\bf M}\cdot {\bf M})^{2},
\label{s1}
\end{eqnarray}  
assuming slowly varying field.
By construction in \ref{s1} the energy densities $f$ and $V$
are clearly rotationally invariant (symmetric).
Higher powers of {\bf M} are neglected, the term
$\alpha_{2}(T)({\bf M}\cdot {\bf M})^{2}$ is kept
since at Curie temperature $\alpha_{1}=\alpha(T-T_{_C}), 
\alpha > 0$ vanishes. The kinetic energy term
$(\partial_{i} {\bf M})^{2}$ is non-negative as
usual, thus to obtain ground-state magnetization 
we must extremize  $V({\bf M})$. This yields
for $T < T_{_C}$ the magnitude of magnetization
\begin{eqnarray}
|{\bf M}| = (-\frac{\alpha_{1}}{2\alpha_{2}})^{1/2} ,
\label{s2}
\end{eqnarray}
which is the ${\em order-parameter}$.

\section{Order-Parameters}

Following the discussion of the previous section
we can easily set-up a GL type of model for
cuprates. The GL type of phenomenological
approach is not uncommon and has been adopted by 
many authors, for example see \cite{tim99}\footnote{The
main differences between reference \cite{tim99} and our
model is that unlike them we don't assume that the
scalar field corresponding to charge is real and that
we deal with couplings of scalar fields with gauge
fields}. The basic physical components of HTSC materials are 
\begin{itemize}
\item{}Charge density.
\item{}Magnetization.
\item{}Superconductivity.
\end{itemize}
Thus generically we may write the free energy density
of such a system in terms of three order parameters,
namely, $\rho$ [charge density], $\phi$ [complex superconducting
order parameter] and $\sigma$ [order parameter for
magnetization]. The free energy density is the sum of
the contributions from each of these order parameters
[each with functional form as in \ref{s1}] plus an interaction
term,
\begin{eqnarray}
f &=& f_{\phi}+ f_{\rho}+ f_{\sigma}+f_{int},\nonumber\\
f_{\phi} &=& (\partial_{i} \phi)^{2}+\alpha_{1}^{\phi} \phi^{*}\phi
-\alpha_{2}^{\phi} (\phi^{*}\phi)^{2},\nonumber\\
f_{\sigma} &=& (\partial_{i} \sigma)^{2}+\alpha_{1}^{\sigma} 
\sigma^{*}\sigma-\alpha_{2}^{\sigma} (\sigma^{*}\sigma)^{2},\nonumber\\
f_{\rho} &=& (\partial_{i} \rho)^{2}+\alpha_{1}^{\rho} 
\rho^{*}\rho-\alpha_{2}^{\rho} (\rho^{*}\rho)^{2},\nonumber\\
f_{int} &=& f_{\phi\sigma}+f_{\phi\rho}+f_{\rho\sigma}.
\label{o1}
\end{eqnarray}  
The simplest choice of the crossed interaction terms
if higher powers than fourth are neglected are of
the form $\alpha_{2}^{\phi\sigma}(\phi^{*}\phi)
(\sigma^{*}\sigma)$ for the $\phi$-$\sigma$ term
and similarly for the rest.

	What we have is essentially three scalar
fields, which represent the order parameters. One
way to proceed is to consider the interaction of scalar
fields with gauge fields and subject this to some
symmetry group. As it is known from particle physics
such considerations lead to some interesting non-perturbative
solutions such as t'Hooft-monopole, Nielsen-Olsen
vortices, instantons and several others. Clearly we have 
many choices for gauge group, but we consider the
most basic ones which can cover the essential physics.
We have to broad choices abelian and non-abelian
groups. The abelian group of direct relevance is
$U(1)$ or $O(2)$ and the one of the simplest
non-abelian choice is $SU(2)$ or $SO(3)$. We
may also consider higher groups such as $SU(N)$ ,
$SO(N)$, $CP^{n-1}$ and others.

The Lagrangian \cite{che82} of gauge-field $A_{\mu}$ 
interacting with a scalar field $\phi$ may be written 
as for abelian case
\begin{eqnarray}
{\cal L} &=& \frac{1}{2}(D_{\mu}\phi)^{\dagger} (D^{\mu}\phi)
+\mu^{2}\phi^{\dagger}\phi-\lambda (\phi^{\dagger}\phi)^{2}
-\frac{1}{4}F_{\mu\nu}F^{\mu\nu},\nonumber\\
D_{\mu}\phi &=& \partial_{\mu} \phi -i g A_{\mu} \phi,\nonumber\\
F_{\mu\nu} &=& \partial_{\mu} A_{\nu}-\partial_{\nu} A_{\mu}, 
\label{o2}
\end{eqnarray}  
and
\begin{eqnarray}
{\cal L} &=& \frac{1}{2}(D_{\mu}^{ij}\phi_{j})^{\dagger} 
(D^{\mu}_{ij}\phi^{j})
+\mu^{2}\phi_{i}^{\dagger}\phi^{i}
-\lambda (\phi^{\dagger}_{i}\phi^{i})^{2}
-\frac{1}{4}F_{\mu\nu}^{i}F^{\mu\nu}_{i},\nonumber\\
D_{\mu}^{ij}\phi_{j} &=& \partial_{\mu} \phi^{i} 
-i g {\cal F}^{ijk} A_{\mu}^{j} \phi_{k},\nonumber\\
F_{\mu\nu}^{i} &=& \partial_{\mu} A_{\nu}^{i}
-\partial_{\nu} A_{\mu}^{i}+g {\cal F}^{ijk}
A_{\mu}^{j}A_{\nu}^{k}, 
\label{o3}
\end{eqnarray} 
for the non-abelian case. The group indices are
denoted by $i,j,k,...$, whereas $\mu,\nu,...$
are the space-time indices. 

	Let us consider SSB in the abelian case.
If the scalar potential is extremized and
the scalar field acquires a vacuum expectation value [VEV]
\begin{eqnarray}
|<0|\phi|0> &=& v/\sqrt{2},\nonumber\\
|\phi|      &=& v/\sqrt{2},\nonumber\\
v           &=& (\mu^2/\lambda)^{1/2}
\label{o4}
\end{eqnarray} 
the symmetry is spontaneously broken. To see this
we first write the $\phi$ field in terms of two real fields 
$\phi_1$ and  $\phi_2$, and choose an arbitrary direction
to achieve SSB, as discussed above, viz 
\begin{eqnarray}
\phi          &=& \frac{1}{\sqrt{2}}(\phi_1+i\phi_2),\nonumber\\
|<0|\phi_1|0> &=& v,\nonumber\\
|<0|\phi_2|0> &=& 0,
\label{o5}
\end{eqnarray} 
After shifting the fields  $\phi_1$ and  $\phi_2$ about
their respective vacuum expectation values we
will end up with the gauge-boson $A_{\mu}$
having acquired a mass $M=gv$, and a scalar
field with mass $m=\sqrt{2}\mu$. The degrees of
freedom [DOF] must correctly add up. We started with
massless gauge boson [2 DOF] and complex $\phi$
field [2 DOF] and ended up with a massive gauge-boson
[3 DOF] and one real scalar field [1 DOF] and
so the DOF correctly add up. The same procedure
easily carries over to the non-abelian case.
For example if we consider the group $SU(2)$, we
would start with three massless gauge fields [6 DOF].
Choosing $\phi$ to be a complex doublet [ 4 DOF]
and assuming SSB we will arrive at 3 massive gauge
bosons [ 9 DOF] and one real scalar field [i.e. Higgs
field] [1 DOF] as one should.

The Lagrangian in \ref{o2} is called the Abelian-gauge
model [AGM]. The connection between the static solution of
AGM and GL model was elegantly pointed out by Nielsen and Olsen
\cite{nie73}. Using the {\em vortex} solution of AGM the connection
with the dual model of strings [Nambu action] was also
pointed out in this work. Thus by looking at particular
solution of a field theory one arrives at a string model.
It is rather unfortunate that although lot of people
in CMT use vortices in context of GL, reference to
\cite{nie73} is almost never made, to our knowledge.
		
	In the above scenario of SSB there is a
physical Higgs, however we may consider other
scenarios where there is {\em no physical Higgs} and
symmetry breaking is non-linearly realized 
\cite{alam98-1} [and references therein].
This approach is same as non-linear sigma
model, which is discussed in \cite{alam00-1}.
For our purposes we can think of non-linear sigma
model as a type of scalar field theory, whose
structure is different from $\phi^{4}$ theory
discussed above and which gives an alternative
description of SSB.

\section{Results}
Our phenomenological model given in Eq.~\ref{o1} consists 
of three scalar fields. Even without including gauge fields it
is non-trivial to solve this model. We thus break it 
into simpler logical parts which we have discussed 
above. For example we can choose to introduce a single 
Abelian gauge field which interacts with charged complex 
field, clearly there are several interesting choices one 
can make. We now outline our results for some of our choices.
\begin{itemize}
\item{}Consider the choice that an Abelian gauge field
interacts with a complex scalar. In the usual manner
we can obtain the string-like vortex solution.
We interpret the string-like solution as charged stripes.
We note that so far the treatment is classical. The interaction
of the other two scalar fields [which represent
superconductivity and magnetism] with these charged
strings can be studied. In a simple sense we can take
the charged scalar field to represent the holons.
The string-like solution \cite{nie73} can be obtained
by assuming that the scalar field approaches a constant
value at large distance. It is straightforward to see that
the minimum of potential is at 
\begin{eqnarray}
|\phi|= \phi_{0}= (\alpha_{1}/(2\alpha_2)^{1/2}.
\label{r1}
\end{eqnarray}
If we expand about this VEV, and denote fluctuations
by $\phi^{'}$, it can be shown that the fluctuations
are defined by a characteristic length $\xi$,
\begin{eqnarray}
\phi^{\prime} &=& e^{-r/\xi}, \nonumber\\
\xi &=& 1/\sqrt(2\alpha_{1}).
\label{r2}
\end{eqnarray}
$\xi$ measures the distance before $|\phi|$ reaches
its asymptotic value, it is the measure of the 
transverse extenstion of the string. Another
relevant parameter is directly related to the 
ratio [r] of the strength of the self-coupling of scalar
field to the interaction strength of scalar 
field ($\alpha_2$) with the gauge field ($e$)
\begin{eqnarray}
r = (2 \alpha_{2}/(e^{2}\alpha_1)^{1/2}.
\label{r3}
\end{eqnarray}
$r$ is a measure over which the electromagnetic field 
is appreciably different from zero. A well-defined string
results if $\xi$ and $r$ have roughly the same magnitude.
Thus the deviation from an exact string structure or
deformed string may be got by choosing suitable values
for these parameters. Such a parameterization is useful
for modeling realistic stripes. Although the string-like 
solution of the AGM is well-known, our interpretation is new, 
namely to interpret the string-like solution as charge stripe. 
We note that Nielsen and Olsen \cite{nie73} found the solution 
by realizing that AGM is the relativistic generalization of 
GL Lagrangian which has a vortex solution as for example 
found by Abrikosov. Yet another new feature is to use
the deviation of the parameters $\xi$ and $r$ from
each other for modelling realistic stripes. 
\item{}
If we choose a non-abelian group such as $SU(2)$, $SU(3)$
or $SO(3)$ we can still obtain a string-like solution
provided we have at least two scalar fields [which are
``vectors'' under the symmetry groups]. The same arguments
go through as discussed for the abelian case. We may use
these solutions to models spin stripes or regions of
spins or spinons. By considering the interactions of
charge and spin stripes we can hope to obtain a phenomenological
model for cuprates and related materials.
\item{}
It is also known [see \cite{nie73}] that in 2+1 dimensions
the simplest non-trivial string-model [or vortex] arises
from the sine-Gordon theory, with Lagrangian, equation
of motion and static solution [with string lying along
the x-axis],
\begin{eqnarray}
{\cal L} &=& \frac{1}{2}(\partial_{\mu}\phi)^{2}
+c \cos (d\phi),\nonumber\\
\Box \phi+cd\sin(d\phi) &=& 0,\nonumber\\
\phi &=& \frac{1}{4}\arctan e^{\sqrt{cd}~y},
\label{r4}
\end{eqnarray}
where $\Box\equiv \partial_{\mu}\partial^{\mu}
 \equiv (1/v^2)\frac{\partial^2}{\partial t^2}
-\nabla^2$ is the familiar wave operator.

We may thus interpret this solution as giving
rise to charge stripes. The above are classical
field theory results. On the other hand it
is known in CMP according to Luttinger liquid 
theory of 1d metals that at large wavelengths
the relevant degrees of freedom are {\em collective
charge and spin oscillations} which are governed
by the quantum sine-Gordon field theories.
Our results are supported by the work of
Zaanen et al. \cite{zaa98}, who have 
used sine-Gordon theory for separation of
spin, charge and string fluctuation.
\end{itemize}
\section{Conclusions}
By considering a set of scalar fields without 
[global theory] or with [local theory] which we
take to define order-parameters of cuprates
we are led naturally led to a  phenomenological
model of strings. These strings and their 
interactions with the relevant degrees of
freedom of systems constitute an interesting
direction to pursue. This approach nicely ties
with and supports our quantum group 
conjecture \cite{alam99-1}, namely that 1d structures 
are relevant to the understanding
of the physics of the cuprates and related materials.
It must be pointed out that almost all
works have used the argument that since quantum groups
were restricted to 1d, it was a problem to use
them. On the contrary we have argued that it is
precisely the opposite, namely 1d systems or
collective excitations are the ones relevant 
\cite{alam99-1} to the physics of the cuprates
and related materials. 
\section*{Acknowledgments}
The Sher Alam's work is supported by the Japan Society for
for Technology [JST].

\end{document}